\documentstyle [12pt]{article}
\title{
   Crystal Collimation Experiment on 70-GeV Proton Accelerator
       }
\author{
A.G.Afonin, \underbar{V.M.Biryukov},
V.N.Chepegin, Yu.A.Chesnokov,\\ V.I.Kotov,
V.I.Terekhov, E.F.Troyanov, Yu.S. Fedotov\\
{\em Institute for High Energy Physics, Protvino, Russia}\\
Yu.M.Ivanov,
{\em Nuclear Physics Institute, St.Petersburg, Russia}\\
W.Scandale,
{\em CERN, Geneva};
M.B.H.Breese,
{\em U. of Surrey, UK}
}
\date{Proc. of 7th ICFA mini-workshop (Lake Como 1999), pp.127-130}
\begin{document}
\maketitle

\begin{abstract}
\normalsize
The first proof-of-principle experiment on "crystal collimation" was
performed with 70-GeV protons on IHEP accelerator.  A bent crystal
installed in the ring as a primary element upstream of a collimator has
reduced the radiation levels downstream in the accelerator by a factor
of two.  The measurements agree with Monte Carlo predictions.
\end{abstract}

\section{Introduction}

Bent-crystal technique is well established for extracting high energy beams
from accelerators. It was successfully applied
at the energies  up to 900 GeV\cite{1},
and simulations were able to predict the results correctly.
Recent experiments at IHEP Protvino\cite{2} have demonstrated
that this technique
can be quite efficient: 50-70\% of the beam have been extracted using
a thin (3-5 mm) {\em Si} channeling crystal with bending of 0.5-1.5 mrad,
with intensity of the extracted 70-GeV beam up to 6$\times$10$^{11}$
protons per spill. At this intensity, no cooling measures were taken and
no reduction in the efficiency observed.
At IHEP Protvino this technique has been routinely used since 1987 to
deliver a 70 GeV beam to particle physics experiments.
One of the IHEP crystals did extract 70 GeV protons over 10 years since 1989
without replacement and without any degradation seen! It was shown in the
experiments at BNL AGS and at CERN SPS that radiation damage in channeling
crystals is sizable only at over (2-4)$\times$10$^{20}$ proton/cm$^2$.

The theory of crystal extraction is based mainly on detailed Monte Carlo
simulations tracking the particles through a curved crystal lattice and the
accelerator environment in a multipass mode. Our code CATCH was successfully
tested in the extraction experiments at CERN, FNAL, and IHEP
in 1992-99\cite{8}.
Monte Carlo predictions, suggesting a "multipass" mode of crystal extraction
where efficiency is dominated by the multiplicity of particle encounters
with a short crystal, have lead to the breakthrough in the extraction
efficiency demonstrated at IHEP Protvino\cite{2}.

Crystal can channel a charged particle
if it comes within so-called critical angle $\theta_c$,
about $\pm$150 $\mu$rad/$\sqrt{pv(GeV)}$ in silicon.
This restricts  crystal efficiency in divergent beams.
However, if a crystal is
installed in a circulating beam,
particle may scatter in inefficient encounters
and have new chances on later turns.
To benefit from the "multi-pass" channeling,
the crystal must be short enough to reduce beam losses
in multiple encounters with it.

It should be promising to apply this bent-crystal technique
for a beam halo scraping in accelerators and storage rings\cite{3,4}.
A bent crystal, serving as a primary element, should bend halo particles
onto a secondary collimator. A demonstration experiment of this kind
was performed at IHEP where for the first time a significant reduction in the
accelerator background was obtained with a bent crystal incorporated
into beam cleaning system\cite{2}.

A crystal collimation system for a gold ion
beam is now being installed at RHIC in collaboration with IHEP\cite{rhic},
and --upon success-- it will serve there on permanent basis.

\section{Crystal Deflector}

Bending a short crystal to be installed in the accelerator
vacuum chamber is not easy.
The first crystal used in the course of our experiment of 1997-1999
was of Si(111) type and performed as a short plate of a big height,
0.5$\times$40$\times$7 mm$^3$ (thickness, height, and length along the beam
direction, respectively).
It was bent transversally with a metal holder which had a hole of 20 mm size
for beam passage, and gave the channeled protons a deflection of 1.7 mrad.
Despite an angular distortion (a "twist") in that design,
encouraging results on beam extraction were obtained in our first run
in December 1997, Figure 1.
The peak extraction efficiency reached about 20\% and
the extracted beam intensity was up to 1.9$\times$10$^{11}$ \cite{iheprep}.
Here and later on in the paper, the extraction efficiency is defined
as the ratio of the extracted beam intensity as measured in the external
beamline to all the beam loss in the accelerator.

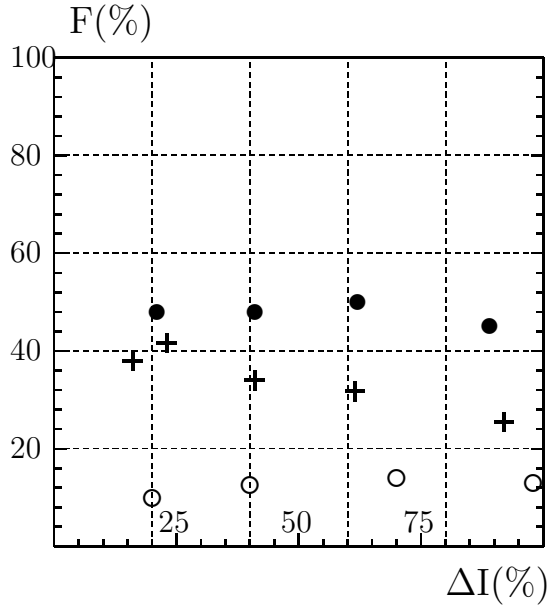
\begin{figure}
	\begin{center}
\setlength{\unitlength}{.65mm}
\begin{picture}(110,120)(0,-3)
\thicklines
\linethickness{.1mm}
\multiput(0,20)(2,0){50}{\line(1,0){1}}
\multiput(0,40)(2,0){50}{\line(1,0){1}}
\multiput(0,60)(2,0){50}{\line(1,0){1}}
\multiput(0,80)(2,0){50}{\line(1,0){1}}
\multiput(20,0)(0,2){50}{\line(0,1){1}}
\multiput(40,0)(0,2){50}{\line(0,1){1}}
\multiput(60,0)(0,2){50}{\line(0,1){1}}
\multiput(80,0)(0,2){50}{\line(0,1){1}}
\linethickness{.4mm}

\put(     90. ,25.5)  {\line(1,0){4}}
\put(     92. ,23.5)  {\line(0,1){4}}
\put(     59.5 ,31.7)  {\line(1,0){4}}
\put(     61.5 ,29.7)  {\line(0,1){4}}
\put(     39. ,34.)  {\line(1,0){4}}
\put(     41. ,32.)  {\line(0,1){4}}
\put(     21. ,41.6)  {\line(1,0){4}}
\put(     23. ,39.6)  {\line(0,1){4}}
\put(     14. ,38.)  {\line(1,0){4}}
\put(     16. ,36.)  {\line(0,1){4}}

\put(     89. ,45.)  {\circle*{3}}
\put(     62. ,50.)  {\circle*{3}}
\put(     41  ,48)  {\circle*{3}}
\put(     21  ,48)  {\circle*{3}}

\put(     98. ,13.)  {\circle {3}}
\put(     70. ,14)  {\circle {3}}
\put(     40. ,12.5)  {\circle {3}}
\put(     20. ,10)  {\circle {3}}

\linethickness{.2mm}
\put(0,0) {\line(1,0){100}}
\put(0,0) {\line(0,1){100}}
\put(0,100) {\line(1,0){100}}
\put(100,0){\line(0,1){100}}
\multiput(25,0)(25,0){3}{\line(0,1){2.5}}
\multiput(5,0)(5,0){20}{\line(0,1){1.4}}
\multiput(0,20)(0,20){4}{\line(1,0){2.5}}
\multiput(0,4)(0,4){25}{\line(1,0){1.4}}
\multiput(100,20)(0,20){4}{\line(-1,0){2.5}}
\multiput(100,4)(0,4){25}{\line(-1,0){1.4}}
\put(24,3){\makebox(1,1)[b]{25}}
\put(49,3){\makebox(1,1)[b]{50}}
\put(74,3){\makebox(1,1)[b]{75}}
\put(-9,20){\makebox(1,.5)[l]{20}}
\put(-9,40){\makebox(1,.5)[l]{40}}
\put(-9,60){\makebox(1,.5)[l]{60}}
\put(-9,80){\makebox(1,.5)[l]{80}}
\put(-9,100){\makebox(1,.5)[l]{100}}

\put(3,105){\large F(\%)}
\put(80,-10){\large $\Delta$I(\%)}

\end{picture}
	\end{center}
	\caption{
	Spill-averaged efficiency of extraction as measured with
	 5-mm crystal 0.65 mrad bent ($\bullet$), December 1998;
	 5-mm crystal 1.5 mrad bent ($+$), March 1998;
	 7-mm twisted crystal 1.7 mrad bent (o), December 1997;
	plotted against the beam fraction taken from the accelerator.
}
	\end{figure}

To further increase the extraction efficiency,
further crystals (without twist)
were made from a monolithic Si piece in a shape of
"O" at the Petersburg Nuclear Physics Institute,
as described in Ref. \cite{plb}.
The crystals Si(110) used in our recent runs had
the length along the beam direction of only 5 mm.
The bent part of the crystal was just 3 mm long,
and the straight ends were 1 mm each.

Such a crystal, with bending angle of 1.5 mrad, was successfully
tested in March 1998 and has shown extraction efficiencies over 40\%
\cite{plb}. In the mean time we have changed the crystal location
in order to use another septum magnet (with partition thickness
of 2.5 mm instead of 8 mm as in the old scheme)
where a smaller bending angle is required from a crystal.
This change was also motivated by the intention to test
even shorter crystals (two of them, 2.5 and 3.0 mm long,
are already undergoing tests).
The crystal used in this location was new, but of the same design
and dimensions as earlier described\cite{plb}.
The bending angle used in this run was 0.65 mrad.

\section{Study of Crystal Work in Slow-Extraction Mode}

Experiments on crystal-assisted slow extraction and scraping are very
similar on the part of crystal component, the only difference being
the target of the channeled deflected beam --- is it an external beamline
or beam absorber. This is why we were able to study the crystal work
first in the conditions of slow extraction where we could measure the
amount and characteristics of the channeled beam more easiely.

The general schematics of beam extraction by a crystal is shown
in Ref.\cite{plb}.
As the small angles of deflection are insufficient for a direct extraction
of the beam from the accelerator,
a crystal served as a primary element in
the existing scheme of slow extraction.
Crystal was placed in straight section 106 of the accelerator
upstream of a septum-magnet of slow-extraction system.
The accuracy of the crystal horizontal and angular
translations was 0.1 mm and 13.5 $\mu$rad, respectively.
The horizontal emittance of the circulating proton beam
was about 2$\pi$ mm$\times$mrad, and
the beam divergence at the crystal location was 0.6 mrad.
A local distortion of the orbit by means of bump windings
in magnets moved the beam slowly toward the crystal.
To obtain a uniform rate of the beam at crystal, a monitor
for close loop operation
based on a photomultiplier with scintillator was used
to automatically adjust the orbit distortion.
We used also function generator to control current in bump windings.

The beam deflection to the septum and its transmission
through the beam line of extraction were supervised with a complex system of
beam diagnostics, including TV system, loss monitors, profilometers,
intensity monitors\cite{plb}.
All the diagnostics devices were firstly tested in
fast-extraction mode and calibrated with beam transformers.
The background conditions were periodically measured
with and without crystal.
According to the measurements, the fraction of background particles
(e.g. elastically scattered protons)
together with the apparatus noise did not exceed 4\% of the
useful signal level. This background was subtracted from
the efficiency figures shown in the paper.
The fraction of the beam directed to the crystal
was defined as the difference
between the measurements of the circulating beam intensity
done with beam transformers
before and after the beam extraction,
with the systematic error of 1\%.
The extraction efficiency
was evaluated in every cycle of acceleration.

\section{Crystal Efficiency}

The accelerator beam intensity during the experiment
was about 1.3$\times$10$^{12}$ protons per cycle.
The fraction of the circulating beam incident on the crystal
$\Delta$I was varied from 20 to 90\%.
The spill duration of the channeled beam in the feedback regime was
on the order of 0.5 s.
The plateau of the IHEP U-70 accelerator magnet cycle is 2 s long
while the overall cycle of the machine is 9.6 s.
Figure 1 shows the efficiency of extraction averaged over the spill,
as measured in our three experiments of 1997-98.
In the last one, the efficiency was about 50\% even when all the
accelerator beam was directed onto the crystal.
The spill-averaged efficiency figures were reproducible
with 1\% accuracy from run to run.
The dependence of the extracted beam intensity
on orientation of the crystal was about the same as in Ref.\cite{plb}
and not shown here.
The highest intensity of the extracted beam, for 1.15$\times$10$^{12}$
protons incident at the crystal in a cycle, was equal to
5.2$\times$10$^{11}$.

As the beam moves radially toward the crystal,
the proton incidence angle drifts at the crystal.
For this reason the extraction efficiency varies in time during the spill,
especially for a large beam fraction used.
The peak extraction efficiency in a spill was always greater
than 60\%.
The absolute extraction efficiency as obtained in
our Monte Carlo simulations agree with the measurements
to accuracy of about 5\% for spill-averaged figures.

\section{Crystal Collimation Experiment}

Bent crystal, situated in the halo of a circulating beam,
can be the primary element in a scraping system,
thus serving as an 'active' collimator.
In this case, the only difference from extraction is that channeled particles
are bent onto a secondary collimator instead of the extraction beamline.
The bent particles are then intercepted (with
sufficiently big impact parameter) at the secondary element and
absorbed there.
\begin{figure}[htb]
\begin{center}
\setlength{\unitlength}{.52mm}
\begin{picture}(150,95)(0,-7)
\thicklines

\put(20,62.2){\large $\diamond$}
\put(45,53.2){\large $\diamond$}
\put(90,40.0){\large $\diamond$}

\put(20,29.5){\large $\bullet$}
\put(45,23.8){\large $\bullet$}
\put(90,17.2){\large $\bullet$}

\put(20,28.4){\bf\large $\circ$}
\put(45,22.4){\bf\large $\circ$}
\put(90,15){\bf\large $\circ$}

\put(20,20){\bf\large $\star$}
\put(45,13.6){\bf\large $\star$}
\put(90,8.8){\bf\large $\star$}

\put(20,7.6){\large $\ast$}
\put(45,5. ){\large $\ast$}
\put(90,2.6){\large $\ast$}

\put(0,0) {\line(1,0){150}}
\put(0,0) {\line(0,1){75}}
\put(0,75) {\line(1,0){150}}
\put(150,0){\line(0,1){75}}
\multiput(0,0)(50,0){3}{\line(0,1){5}}
\multiput(0,0)(5,0){30}{\line(0,1){2}}
\multiput(0,10)(0,10){7}{\line(1,0){3}}
\put(0,-6){\makebox(1,1)[b]{ 0}}
\put(50,-6){\makebox(1,1)[b]{5}}
\put(100,-6){\makebox(1,1)[b]{10}}
\put(150,-6){\makebox(1,1)[b]{15}}
\put(-6,0){\makebox(1,.5)[l]{0}}
\put(-6,10){\makebox(1,.5)[l]{1}}
\put(-6,20){\makebox(1,.5)[l]{2}}
\put(-6,30){\makebox(1,.5)[l]{3}}
\put(-6,40){\makebox(1,.5)[l]{4}}
\put(-6,50){\makebox(1,.5)[l]{5}}
\put(-6,60){\makebox(1,.5)[l]{6}}
\put(-6,70){\makebox(1,.5)[l]{7}}

\put(5,80){I (rel.)}
\put(20,-14){Distance from collimator (m)}

\end{picture}
\end{center}
\caption{
Radiation levels as monitored at three places along the ring
in the vicinity of FEP, for different cases (bottom up):
$\ast$ - beam kicked onto absorber by a kicker magnet;
$\star$ - aligned crystal as primary;
$\circ$ - FEP works as primary;
$\bullet$ - misaligned crystal as primary;
$\diamond$ - Si target downstream of FEP is primary.
}
  \label{col0}
\end{figure}
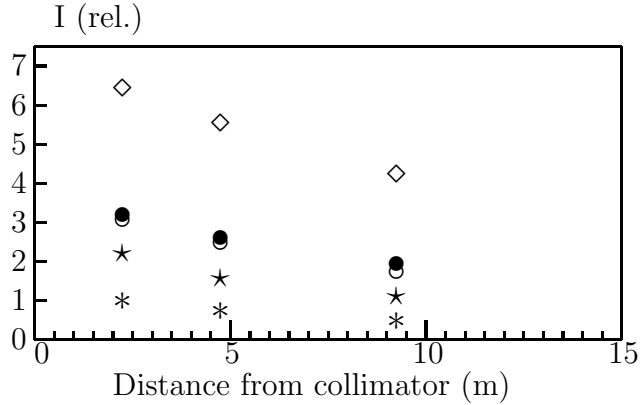

We have performed the first demonstration experiment
on crystal-assisted collimation.
A bent crystal, with the same dimensions as the extraction crystals
described above and with bending angle of 1 mrad, was positioned
upstream of a secondary collimator
(stainless steel absorber 4 cm wide, 18 cm high, 250 cm long) "FEP"
and closer to the beam in the horizontal plane.
As the horizontal betatron tune is 9.73 in our accelerator,
it was most convenient to intercept the bent beam at FEP
not immediately on the first turn, but after 3 turns in the accelerator.
In this case the deflection angle of 1 mrad transforms into
more than 20 mm horizontal offset, and so the bent beam enters
the FEP collimator at some $\sim$15 mm from the FEP edge.
The optimal horizontal position of the crystal w.r.t. the FEP edge
was found to be $\sim$10 mm.

Radiation levels were monitored at three places along the ring
in the vicinity of FEP, from $\sim$2 to $\sim$10 meters downstream
of the backward edge of the collimator.
Several different cases have been studied.
\begin{itemize}
\item
The whole accelerator beam was kicked into the middle of the FEP face
by a kicker magnet. That's an ideal case for a beam interception and
absorption, so the resulting radiation levels (nonzero due to escape
of some primary and secondary particles from the FEP body)
can be considered as a kind of pedestal for the results obtained then
with several actual scraping methods. These lowest levels are shown
in Figure 2 by ($\ast$) marks.
\item
When FEP was a primary element scraping the beam halo continously,
the halo particles were entering FEP very close to its edge (at sub-micron
depths) so the escape of particles from FEP body
because of outscattering was very important. This resulted in higher
radiation levels ($\circ$) as shown in Figure 2.
\item
A bent silicon crystal was introduced then about 60 cm upstream
of the forward edge of FEP. Crystal served as a primary element
of the scraping system, being closer to the circulating beam than FEP,
with the offset of about 5-15 mm in the radial plane.
When the crystal was misaligned, it was acting as an amorphous target
scattering particles. The collimator downstream could intercept some
of the scattered particles. The radiation levels measured ($\bullet$)
in this setting were not so different from the preceding case of direct
(by FEP) scraping of the beam halo.
\item
When the crystal was aligned to the best angle w.r.t. the incident beam,
a substantial number of halo particles was channeled and deflected
into the depth of FEP for best absorption. The monitored radiation
levels with aligned crystal serving as primary element are shown
($\star$) in Figure 2. One can conclude that about one half of the
halo was extracted and forwarded to a safe place (i.e. the middle of FEP
face) for absorption, reducing the radiation background in the ring
correspondingly.
\item
Finally, another case studied was a silicon target (amorphous)
positioned downstream of FEP as a primary element.
In this case the machine was not protected from the scattered particles
originating in the target, so the radiation levels achieved
($\diamond$) were the highest.
\end{itemize}
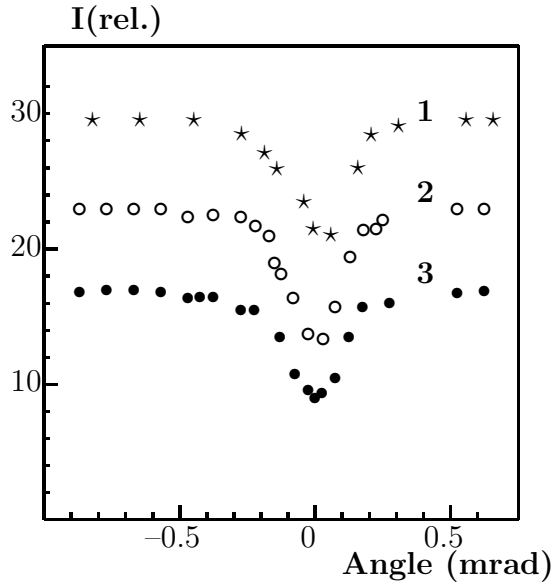
\begin{figure}[htb]
\begin{center}
\setlength{\unitlength}{1.8mm}
\begin{picture}(35,42)(0,-4)
\thicklines
\linethickness{.3mm}

\put(    27.5, 29.5)  {\bf 1}
\put(    32.6, 29)  {$\star$}
\put(    30.6, 29)  {$\star$}
\put(    25.6, 28.6)  {$\star$}
\put(    23.6, 27.9)  {$\star$}
\put(    22.6, 25.5)  {$\star$}
\put(    20.6, 20.5)  {$\star$}
\put(    19.3, 21)  {$\star$}
\put(    18.6, 23)  {$\star$}
\put(    16.6, 25.4)  {$\star$}
\put(    15.7, 26.6)  {$\star$}
\put(    14, 28)  {$\star$}
\put(    10.5, 29)  {$\star$}
\put(    6.5, 29)  {$\star$}
\put(    3, 29)  {$\star$}

\put(    27.5, 17.5)  {\bf 3}
\put(    32.5, 16.9)  {\circle*{.8}}
\put(    30.5, 16.75)  {\circle*{.8}}
\put(    25.5, 16)  {\circle*{.8}}
\put(    23.5, 15.7)  {\circle*{.8}}
\put(    22.5, 13.5)  {\circle*{.8}}
\put(    21.5, 10.5)  {\circle*{.8}}
\put(    20.5, 9.4)  {\circle*{.8}}
\put(    20, 9.)  {\circle*{.8}}
\put(    19.5, 9.6)  {\circle*{.8}}
\put(    18.5, 10.8)  {\circle*{.8}}
\put(    17.4, 13.5)  {\circle*{.8}}
\put(    15.5, 15.5)  {\circle*{.8}}
\put(    14.5, 15.5)  {\circle*{.8}}
\put(    12.5, 16.5)  {\circle*{.8}}
\put(    11.5, 16.5)  {\circle*{.8}}
\put(    10.6, 16.4)  {\circle*{.8}}
\put(    8.6, 16.8)  {\circle*{.8}}
\put(    6.6, 17)  {\circle*{.8}}
\put(    4.6, 17)  {\circle*{.8}}
\put(    2.6, 16.8)  {\circle*{.8}}

\put(    27.5, 23.5)  {\bf 2}
\put(    32.5, 23)  {\circle{.8}}
\put(    30.5, 23)  {\circle{.8}}
\put(    25, 22.2)  {\circle{.8}}
\put(    24.5, 21.5)  {\circle{.8}}
\put(    23.6, 21.4)  {\circle{.8}}
\put(    22.6, 19.4)  {\circle{.8}}
\put(    21.5, 15.7)  {\circle{.8}}
\put(    20.6, 13.4)  {\circle{.8}}
\put(    19.5, 13.7)  {\circle{.8}}
\put(    18.4, 16.4)  {\circle{.8}}
\put(    17.5, 18.2)  {\circle{.8}}
\put(    17, 19)  {\circle{.8}}
\put(    16.6, 21)  {\circle{.8}}
\put(    15.6, 21.7)  {\circle{.8}}
\put(    14.5, 22.4)  {\circle{.8}}
\put(    12.5, 22.5)  {\circle{.8}}
\put(    10.6, 22.4)  {\circle{.8}}
\put(    8.6, 23)  {\circle{.8}}
\put(    6.6, 23)  {\circle{.8}}
\put(    4.6, 23)  {\circle{.8}}
\put(    2.6, 23)  {\circle{.8}}

\put(0,0) {\line(1,0){35}}
\put(0,0) {\line(0,1){35}}
\put(0,35) {\line(1,0){35}}
\put(35,0){\line(0,1){35}}
\multiput(0,0)(2,0){17}{\line(0,1){.5}}
\multiput(0,0)(10,0){4}{\line(0,1){1}}
\multiput(0,0)(0,10){3}{\line(1,0){1}}
\multiput(0,0)(0,2){17}{\line(1,0){.4}}
\put(9,-2){\makebox(1,1)[b]{--0.5}}
\put(19,-2){\makebox(1,1)[b]{0}}
\put(29,-2){\makebox(1,1)[b]{0.5}}
\put(-2.5,30){\makebox(1,.5)[l]{30}}
\put(-2.5,20){\makebox(1,.5)[l]{20}}
\put(-2.5,10){\makebox(1,.5)[l]{10}}

\put(2,36.3){\bf I(rel.)}
\put(22,-4){\bf Angle (mrad)}

\end{picture}
\end{center}
\caption{
Measured irradiation in detectors 1, 2, 3 as function of crystal angle.
}
  \label{col1}
\end{figure}

Figure 3   shows how the radiation level depends on the angular
alignment of the crystal.
At the best crystal angle, preferable for channeling,
the radiation levels decrease by up to factor of $\sim$two in
the  places of monitoring.
This is explained by the fact that $\sim$50\% of the incident beam
is channeled by the crystal and deflected to the depth of FEP
where absorbed.
In the case when crystal was out and the beam was scraped
directly by FEP,
the radiation at the monitors was at about
the same level as in the case of disaligned crystal.
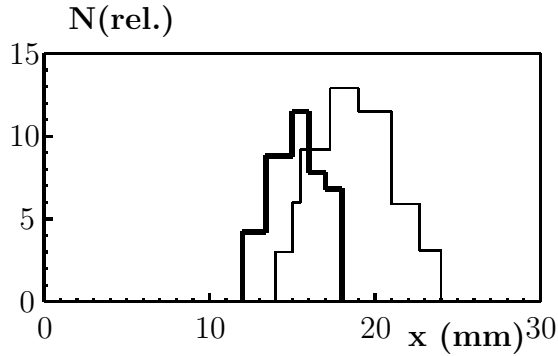
\begin{figure}[htb]
\begin{center}
\setlength{\unitlength}{.22mm}
\begin{picture}(300,197)(0,-6)
\thicklines

\put(0,0) {\line(1,0){300}}
\put(0,0) {\line(0,1){150}}
\put(0,150) {\line(1,0){300}}
\put(300,0){\line(0,1){150}}
\multiput(0,0)(100,0){3}{\line(0,1){5}}
\multiput(0,0)(10,0){30}{\line(0,1){2}}
\multiput(0,50)(0,50){3}{\line(1,0){5}}
\multiput(0,0)(0,10){15}{\line(1,0){2}}
\put(0,-22){\makebox(1,1)[b]{0}}
\put(100,-22){\makebox(1,1)[b]{10}}
\put(200,-22){\makebox(1,1)[b]{20}}
\put(300,-22){\makebox(1,1)[b]{30}}
\put(-21,0){\makebox(1,.5)[l]{ 0}}
\put(-21,50){\makebox(1,.5)[l]{ 5}}
\put(-21,100){\makebox(1,.5)[l]{10}}
\put(-21,150){\makebox(1,.5)[l]{15}}

\put(15,165){\bf N(rel.)}
\put(220,-27){\bf x (mm)}

\linethickness{.6mm}

\put(    180.,68) {\line(0,-1){68}}
\put(    170.,68) {\line(1,0){10}}
\put(    170.,78) {\line(0,-1){10}}
\put(    160.,78) {\line(1,0){10}}
\put(    160.,115) {\line(0,-1){37}}
\put(    150.,115) {\line(1,0){10}}
\put(    150.,88) {\line(0,1){27}}
\put(    134.,88) {\line(1,0){16}}
\put(    134.,42) {\line(0,1){46}}
\put(    120.,42) {\line(1,0){14}}
\put(    120.,0) {\line(0,1){42}}

\linethickness{.2mm}
\put(    240.,31)  {\line(0,-1){31}}
\put(    227.,31)  {\line(1,0){13}}
\put(    227.,59)  {\line(0,-1){28}}
\put(    210.,59)  {\line(1,0){17}}
\put(    210.,115)  {\line(0,-1){56}}
\put(    190.,115)  {\line(1,0){20}}
\put(    190.,129)  {\line(0,-1){14}}
\put(    173.,129)  {\line(1,0){17}}
\put(    173.,92)  {\line(0,1){37}}
\put(    155.,92)  {\line(1,0){18}}
\put(    155.,60)  {\line(0,1){32}}
\put(    150.,60)  {\line(1,0){5}}
\put(    150.,30)  {\line(0,1){30}}
\put(    140.,30)  {\line(1,0){10}}
\put(    140.,0)  {\line(0,1){30}}

\end{picture}
\end{center}
\caption{
Profiles measured at FEP entry face:
the channeled beam (thick line)
and the beam (thin line) deflected by kicker magnet.
}
  \label{col2}
\end{figure}

We were able to check the crystal efficiency figure by alternative
means, measuring the profile and intensity of the particles
incident at the FEP entry face. The channeled beam had a narrow
profile and was well distanced from the FEP edge,
as shows Figure 4  where this profile is shown in comparison
with the profile of the accelerator beam deflected onto FEP by a kicker
magnet. From comparison of the two profiles, from crystal and from kicker,
we again derived the crystal efficiency, which was found to be
about 50\%, in agreement with the radiation monitoring figures
and with the earlier shown figures of extraction efficiency
with crystal in straight section 106.

\section{Conclusions}

The crystal-assisted method of beam steering (for scraping or slow
extraction) demonstrates
peak efficiencies in the order of 60-70\%
and shows reliable, reproducible and predictable work.
Crystal can channel at least 5-6$\times$10$^{11}$ ppp
with no cooling measures taken and no degradation seen.

In our experiment
this technique was for the first time demonstrated for
scraping of the beam halo. Such application has been studied
by computer simulation for several machines,
notably RHIC \cite{rhic} and Tevatron \cite{tev}.
We have shown that radiation levels in accelerator can be
significantly decreased by means of channeling crystal incorporated
into beam cleaning system as a primary element.

We continue tests with crystals as short as down to 1 mm,
where Monte Carlo predicts 80-90\% efficiency of steering.
We study different techniques to prepare bent crystal lattices with
required size, one of the most interesting approaches is described
in Ref.\cite{breese}.

\section{Acknowledgements}

The author thanks very much the conference Organizers,
Nikolai Mokhov and Weiren Chou, for kind hospitality and support.

\end{document}